# Computational Intelligence: are you crazy? Since when has intelligence become computational?

Emanuel Diamant
VIDIA-mant,
Kiriat Ono 55100 Israel
Email: {emanl.245@gmail.com}

*Abstract*— Computational Intelligence is a dead-end attempt to recreate human-like intelligence in a computing machine. The goal is unattainable because the means chosen for its accomplishment are mutually inconsistent and contradictory: "Computational" implies data processing ability while "Intelligence" implies the ability to process information. In the research community, there is a lack of interest in data versus information divergence. The cause of this indifference is the Shannon's Information theory, which has dominated the scientific community since the early 1950s. However, today it is clear that Shannon's theory is applicable only to a specific case of data communication and is inapplicable to the majority of other occasions, where information about semantic properties of a message must be taken into account. The paper will try to explain the devastating results of overlooking some of these very important issues – what is intelligence, what is semantic information, how they are interrelated and what happens when the relationship is disregarded.

## I. Introduction

Computational Intelligence (CI) is a branch of Artificial Intelligence (AI), and it seizes its goals from the aspirations specific to the whole family – to recreate human-like intelligence in a human-made machine. What distinguish it from other family members is the means that are chosen to reach the goal – a computer, a computational approach. At the mid of the past century, that was a prevalent and a natural paradigm for the scientific practice – the computer has just invaded our lives, "brain as a computer" came to be as a popular metaphor of this time, and all around the world has at once become computable: Computational chemistry, Computational ecology, Computational genomics, Computational neuroscience, Computational linguistics, Computational intelligence, and so on.

The term Intelligence, as it was already explained, was inherited from the parent's family name, from the Artificial Intelligence. As in the parent's case, the term is ambiguous, blurred and doubtful. AI was invented at about the same time (the time of computer dawn), in the summer of 1956, by four brilliant scientists: J. McCarthy, M.L. Minsky, N. Rochester, and C.E. Shannon. Despite the prominence of the "founding fathers", they have failed to assess the complexity of the task.

It was assumed that the best-known manifestation of intelligence is human intelligence; therefore, AI's aim was defined as human intelligence replication. It was also assumed that, because the brain is the core of intelligence and the brain is busy with information processing, intelligence should be defined as a product of information processing. Hence, one of the AI's destinations was assigned as information processing. And this assignment has been later inherited by the CI.

To the end of the century, it has become clear that "computational" and "intelligence" are terms that belong to different fields of studies, and the two are incompatible. "Computational" implies data processing and "intelligence" implies information processing. The two processing paradigms are clashing. Although fifty years ago the difference between them was not perceived clear enough. The terms "data" and "information" even today continue to be used interchangeably and in a transposable fashion. The reason for this is the Shannon's "Mathematical Theory of Communication", [1], and the Information Theory embedded in it. For a long time, during all the second half of the past century, Shannon's Information Theory was the dominant research paradigm of the scientific community. The original aim of the theory was to solve a purely technical problem: to increase the performance of a communication system. In his theory, Shannon defines information in terms of signal's statistical properties and the uncertainty of receiving a particular signal among those that are possible. The theory has explicitly linked information notion with data and set aside any discussion about signal's value or meaning.

In the year 1949, Shannon wrote: "These semantic aspects of communication are irrelevant to the engineering problem… It is important to emphasize, at the start, that we are not concerned with the meaning or the truth of messages; semantics lies outside the scope of mathematical information theory", [2].

However, in contemporary science, semantic aspects of a message are of a paramount importance. But fascinated with the achievements of Information theory in the communication domain, various scientific communities were eager to apply it almost in every other research field. That forced Shannon to issue an additional warning (in 1956): "In short, information theory is currently partaking of a somewhat heady draught of general popularity. It will be all too easy for our somewhat artificial prosperity to collapse overnight when it is realized that the use of a few exciting words like information, entropy, redundancy, do not solve all our problems", [3].

Yet the mainstream sciences continue to ignore Shannon's warnings. Therefore, even today, the interrelations between "information" and "data", "information" and "semantics", "semantics" and "knowledge" remain undefined, blurred and intuitive (due to the heritage of Information Theory).

It must be mentioned (in this regard) that the first attempt to clarify the relations between "information" and "semantics" was made about 60 years ago by Yehoshua Bar-Hillel and Rudolf Carnap, [4]. As to my knowledge, they were the first who coined the term "Semantic Information". They have sincerely believed that such a merging can be possible: "Prevailing theory of communication (or transmission of information) deliberately neglects the semantic aspects of communication, i. e., the meaning of the messages… Instead of dealing with the information carried by letters, sound waves, and the like, we may talk about the information carried by the sentence", [4].

However, they were not successful in their attempt to unite the mathematical theory of information and semantics. The mainstream thinking at that time was determined by The Mathematical Theory of Communication, which does not distinguish between data and information. By today's standards, the distinction between data and information is irrelevant and meaningless. For this reason, the issue of information handling (specifically semantic information handling) remains neglected and unsettled. Therefore, it will be our duty to try to address these challenging issues (data – information – semantic information – knowledge interrelations) and to show how these interrelations affect CI development.

## II. WHO IS WHO: DATA, INFORMATION, SEMANTICS

As it was said, Shannon defines information as the entropy of a discrete set of probabilities, as an opportunity to reduce uncertainty of a received data transfer. My approach to information relies on the Kolmogorov's approach (to the subject) [5].

A slightly modified and extended version of Kolmogorov's description sounds today (in my words) like this: **"Information is a linguistic description of structures observable in a given data set".**

To make the scrutiny into this definition more palpable I propose a digital image to be considered as a given data set.

A digital image is a two-dimensional set of data elements called picture elements or pixels. In an image, pixels are distributed not randomly, but, due to the similarity in their physical properties, they are naturally grouped into some clusters or clumps. I propose to call these clusters **primary or physical data structures**.

In the eyes of an external observer, the primary data structures are further arranged into more larger and complex agglomerations, which I propose to call **secondary data structures**. These secondary structures reflect human observer aptitude to the arrangement of the primary data structures, and therefore they could be called **meaningful or semantic data structures**. While formation of primary (physical) data structures is guided by objective (natural, physical) properties of the data, the ensuing formation of secondary (semantic) data structures is a subjective process guided by human conventions and habits.

As it was said, **Description of structures observable in a data set should be called "Information".** In this regard, two types of information must be distinguished – **Physical Information and Semantic Information**. They are both language-based descriptions; however, physical information can be described with a variety of languages (recall that mathematics is also a language), while semantic information can be described only by means of natural human language. (More details on the subject you can find in [6]).

Those, who will go and look in [6], would find out that every information description is a top-down evolving coarse-to-fine hierarchy of descriptions that represent various levels of description complexity (various levels of description details). Physical information hierarchy is located at the lowest level of the semantic hierarchy. The process of sensor data interpretation is reified as a process of physical information extraction from the input data, followed by an attempt to associate this physical information (about the input data) with physical information already retained at the lowest level of the semantic hierarchy. If such an association is achieved, the input physical information becomes related (via the physical information retained in the system) with a relevant linguistic term, with a word that places the physical information in the context of a phrase, which provides the semantic interpretation of it. In such a way, the input physical information becomes named with an appropriate linguistic label and framed into a suitable linguistic phrase (and further – in a story, a tale, a narrative), which provides the desired meaning for the input physical information.

The segregation between physical and semantic information is the most essential insight about the nature of information. Another insight is that, because of the subjective nature of semantic information, its creation cannot be formalized. Semantic information hierarchy, thus, cannot be learned and has to be provided to the system always from the outside, always as a gift, a grant, an offering. The next important outcome from the definition given above is the understanding that information descriptions are always reified as a string of words, a piece of text, a narrative.

Bearing in mind all these new peculiarities, we can proceed to further revision of information processing implications for the research into CI topics.

## III. IT'S TIME TO CHANGE YOUR MIND

At the dawn of the AI era, its founding fathers have failed to define the notion of the term "Intelligence" but the common wisdom had led the society to a belief that, because the brain is the core of intelligence and the brain is busy with information processing, intelligence has to be considered as a product of information processing. At that time, information was understood in Shannon's sense, that is, unseparated from data and therefore the construction "Computational

Intelligence" has looked pretty justified – data based (Computational) information processing (Intelligence).

However, historically, the suitability of the term "Computational Intelligence" was challenged very early. In the mid-nineties of the past century, Prof. Lotfi Zadeh has introduced a paradigm, which he dubbed as "Computing with Words" (CWW) paradigm, [7]. CWW was proposed as "a system of computation which offers an important capability that traditional systems do not have—a **capability to compute with information described in natural language",** [8]. ("Information described in natural language" fits exactly my definition of semantic information, but two decades ago, this notion of semantic information has not been existing yet).

It is worth to be mentioned that the "computing" component of the paradigm name has not arouse any objections from the side of Prof. Zadeh, although he was aware about its inappropriateness: "Computing, in its usual sense, is centered on manipulation of numbers and symbols. In contrast, computing with words, or CW for short, is a methodology in which the objects of computation are words and propositions drawn from a natural language", [9].

To avoid any blames of misrepresentation of the CWW principles, I will keep on to exploit extensive citations drawn from the Prof. Zadeh's seminal papers. And that is what you can learn from them: "Computing with words is inspired by the remarkable human capability to perform a wide variety of physical and mental tasks without any measurements and any computations. Underlying this remarkable capability is the brain's crucial ability to manipulate perceptions − perceptions of distance, size, weight, color, speed, time, direction, force, number, truth, likelihood and other characteristics of physical and mental objects. Manipulation of perceptions plays a key role in human recognition, decision and execution processes. As a methodology, computing with words provides a foundation for a computational theory of perceptions… A basic difference between perceptions and measurements is that, in general, measurements are crisp whereas perceptions are fuzzy..." [9].

And again: Computing with words assumes that "computers would be activated by words, which would be converted into a mathematical representation using fuzzy sets (FSs), and that these FSs would be mapped by means of a CWW engine into some other FS, after which the latter would be converted back into a word "[10].

I dare to say that, despite of his intelligence and intuition, Zadeh's way of thinking was plagued by Shannon's perception of information – he does not distinguish between data processing and information processing. As it follows from my definition of information (given above in this paper), "mathematical representation" and "fuzzy sets" usage could be applied only to data structures observable in a data set. Essentially, "mathematical representation" implies physical information processing, although the term "physical information" is unknown to the public at that time. That is the reason why the term "computing" is used in all the cases debated above. Using "words" in the CWW paradigm name implies that what is really kept in mind is the semantic information associated with these words. However, the notion of "semantic information" also does not exist at that time.

There is a widespread assumption that CI designers have reached remarkable achievements "developing cognitive algorithms for engineering applications" based on artificial neural networks, fuzzy logic systems, evolutionary learning algorithms. That is a false assumption about CI achievements. "Developing cognitive algorithms" is a misunderstanding like Cognitive Computing, Computational Intelligence or Computing with Words. The term "Algorithmic information" was introduced by Ray Solomonoff (at a Conference at Caltech in 1960) and further developed by Gregory Chaitin (in 1965). According to Gregory Chaitin, the theory is "the result of putting Shannon's information theory and Turing's computability theory into a cocktail shaker and shaking vigorously", [11]. In the light of what is advocated in this paper that means that the term "algorithmic" can be applied only to cases considering some sort of data processing (physical information, e.g.). Intelligence, cognition and consciousness, as well as all other derivatives of semantic information could not be drawn from data processing.

I accept Prof. Zadeh's insight that "computers would be activated by words", but I don't know yet how it could be done. Semantic information is a string of words, a piece of text. Therefore, semantic information processing indisputably presumes text processing. Contemporary computers are not appropriate for such a task, because today's computers are data processing machines only. It will be our duty to overcome this challenge.

IV. CONCLUSIONS

Computational Intelligence is a viable field of scientific exploration aimed to resolve the most critical problem of our time – to enable meaningful handling of huge volumes of data inundating today our surroundings. It is in line with some other modern technological undertakings targeted on the same goal – the meaningful handling of big data volumes. These undertakings are well known – Cognitive Computing, Computing with Words, Artificial and Machine Intelligence. All they are supposed to serve as main research tools in specific application fields like biology, neuroscience, linguistics, and so on.

In this paper, I have tried to explain why the terms "Computational Intelligence", "Cognitive Computing", "Computing with Words", and "Machine Intelligence" are all wrong and misleading. Essentially, they all are oxymorons – a figure of speech that connects two contradictory terms, like, for example, "exact approximation" or "certainly possible". The term "Computational" was inherited from the early 50s of the past century, when computers have invaded our lives and everything around us has become computable. All branches of science, thus, have become computational: Computational biology, Computational genomics, Computational ecology, Computational linguistics, and so on. Brain sciences were not an exception in this regard – Computational neuroscience, Computational intelligence, and alike.

However, in the past decades, the situation changes substantially – we witness a paradigm shift from a data processing (computational) approach to an information processing (cognitive) approach. ("Cognitive" here implies "capable of information processing"). Not in one day, but gradually, the "Computational" sciences are becoming replaced with "Cognitive" sciences – Cognitive biology, Cognitive neuroscience, Cognitive endocrinology, Cognitive linguistics, and so on.

This tendency is hampered by a lack of understanding about what is "information processing", and, subsequently, comprehension of "what is information". I hope that the humble explanations given in this paper will help the people to find a better way to handle meaningfully the storming streams of big data deluge.